\documentclass[english]{revtex4}
\usepackage[T1]{fontenc}
\usepackage[latin9]{inputenc}
\usepackage{float}
\usepackage{amsmath}
\usepackage{graphicx}
\usepackage{amssymb}

\makeatletter
\@ifundefined{textcolor}{}
{
 \definecolor{BLACK}{gray}{0}
 \definecolor{WHITE}{gray}{1}
 \definecolor{RED}{rgb}{1,0,0}
 \definecolor{GREEN}{rgb}{0,1,0}
 \definecolor{BLUE}{rgb}{0,0,1}
 \definecolor{CYAN}{cmyk}{1,0,0,0}
 \definecolor{MAGENTA}{cmyk}{0,1,0,0}
 \definecolor{YELLOW}{cmyk}{0,0,1,0}
 }

\usepackage{float}

\makeatletter
\@ifundefined{textcolor}{}
{
 \definecolor{BLACK}{gray}{0}
 \definecolor{WHITE}{gray}{1}
 \definecolor{RED}{rgb}{1,0,0}
 \definecolor{GREEN}{rgb}{0,1,0}
 \definecolor{BLUE}{rgb}{0,0,1}
 \definecolor{CYAN}{cmyk}{1,0,0,0}
 \definecolor{MAGENTA}{cmyk}{0,1,0,0}
 \definecolor{YELLOW}{cmyk}{0,0,1,0}
 }

\makeatother

\usepackage{babel}

\begin{document}

\title{Planetary systems based on a quantum-like model}

\author{N. Poveda T.}
\email{nicanor.poveda@uptc.edu.co}
\affiliation{Universidad Pedagógica y Tecnológica de Colombia, Grupo de Astrofísica
y Cosmología}

\author{N. Vera-Villamizar.}
\email{nelson.vera@uptc.edu.co}
\affiliation{Universidad Pedagógica y Tecnológica de Colombia, Grupo de Astrofísica
y Cosmología}

\author{N. Y. Buitrago C.}
\email{nidia.buitrago@uptc.edu.co}
\affiliation{Universidad Pedagógica y Tecnológica de Colombia, Grupo de Astrofísica
y Cosmología}

\begin{abstract}
Planetary systems have their origin in the gravitational collapse
of a cloud of gas and dust. Through a process of accretion, is formed
a massive star and a disk of planetesimals orbiting the star. Using
a formalism analogous to quantum mechanics (quantum-like model), the
star-planetesimal system is described and the flow quantizing the
gravitational field theoretical model parameters are obtained. Goodness
of fit (chi-square) of the observed data with model quantum-like,
to the solar system, satellites, exoplanets and protoplanetary disk
around HL Tauri is determined. Shows that the radius, eccentricity,
energy, angular momentum and orbital inclination of planetary objects
formed take discrete values depending only on the mass star.
\end{abstract}

\maketitle

\section{Introduction}

In the solar system the distance of the planets from the Sun follow
a simple geometric progression, known as the Titius-Bode law: $r=0.4+0.3\times2^{n}\mbox{ AU}$
with $n=-\infty,\ 0,1,\ldots,7$ \citep{Lynch,Neslusan}. However,
better agreement using the Bohr equation is obtained: $r=0.0425\times n^{2}\mbox{AU}$,
with $n=3,4,5,6,8,11,15,21,27$ \citep{Caswell,Penniston,Barnothy1,Barnothy2}.
Comparisons of the Titius-Bode law with the equation of Bohr raised
the issue of the applicability of certain principles of quantum mechanical
in systems orbital of the astronomy \citep{Corliss}. To interpret
Bohr's equation within the framework of classical mechanics, the quantization
rule of Bohr-Sommerfeld was used \citep{Agnese-Festa}. These basic
ideas were extended to the Jovian satellites \citep{Rubcic1,Rubcic3,Hermann}
and extrasolar planets \citep{Rubcic2,Rubcic4}. Finally, it has attempted
to describe planetary systems with the Schrödinger equation \citep{Reinisch,Oliveira,Smarandache,Nie}.

In quantum mechanics, Planck's constant is extremely small, $h\sim10^{-34}\mbox{J.s}$
and the ratio de Broglie wavelength is inversely proportional to momentum.
For a massive object length de Broglie wave is very small, and consequently
the wave behavior of a macroscopic object is undetectable, making
quantum mechanics reduces to classical mechanics. Thus quantum mechanics
is a theory developed to explain phenomena only a microscopic scale.

Planetary systems originate from a nebula which collapses to form
a star and a disk of planetesimals orbiting the star. The trajectory
of the planetesimals is determined by the principle of stationary
action, $\delta S=0$. However, planetesimal is disturbed by the presence
of the other, oscillating around the classical trayectory; this causes
for a closed and stable trayectory, the action is quantized; that
is, the action is an integer multiple of a quantum of action, $S=nh_{s}$
(see Appendix A). The quantum of action, $h_{s}$ (which plays a role
equivalent to Planck's constant $h$) is a free parameter to be determined
and depends on the physical system in question. Considering the flux
quantization of the gravitational field, the value of $h_{s}$ is
determined. For a massive object: $h_{s}\gg h$, this allows to describe
a macroscopic system with the formalism of quantum mechanics (quantum-like
model). This means that the formalism of quantum mechanics is applicable
to any system that quantize the action. One could use the Bohmian
interpretation to give meaning to the formalism.

This implies that planetary systems (planets-Sun, satellites-planet,
exoplanets-star, exoplanets-pulsar) must quantize not only the orbital
radius as demonstrated, but also eccentricity, orbital inclination,
angular momentum and energy. Through the goodness of fit (chi-square)
shows that there is a very good agreement between the observed data
and the quantum-like model, although the formulation is based only
on the interaction between two particles.

\section{FLUX QUANTIZATION OF THE GRAVITATIONAL FIELD}

In a microscopic system, the electric potential energy between a proton
and an electron, is given by, $U_{e}=-e^{2}/4\pi\epsilon_{o}$, and
an equivalent macroscopic system: the gravitational potential energy
between a star and a planetesimal, is given by, $U_{g}=-GM_{s}m/r$.
A relationship of proportionality between the electric potential energy
and the quantum of action $h$ (Planck's constant), is assumed: $U_{g}\sim h$,
also between the gravitational potential energy and $h_{s}$ parameter:
$U_{g}\sim h_{s}$, which plays the same role of Planck's constant,
in a macroscopic system. We can establish the relationship between
these two energies, to obtain a dimensionless independent amount of
$r$, and normalize the macroscopic parameter on a unit, the Planck
constant: $U_{g}/U_{e}\sim\hbar_{s}/\hbar=\widetilde{\hbar}$.

The flow of the gravitational field generated by the mass of a star,
through a spherical gaussian surface, is given by: $\Phi_{g}=4\pi GM_{s}$
and the electric field flux generated by a proton is: $\Phi_{e}=e/\epsilon_{o}$.
If we take the relation between the intensity of the gravitational
and electromagnetic force in terms of their flows, $\widetilde{\Phi}=m\Phi_{g}/e\Phi_{e}$,
we have, $\widetilde{\Phi}\sim\widetilde{\hbar}$. It is considered
$\widetilde{\Phi}$, as flow of the quantum of action per unit solid
angle: $\widetilde{\Phi}=\widetilde{\hbar}/\Omega$ ($\Omega$ acts
as the proportionality constant), we obtain 
\[
\frac{4\pi GM_{s}\hbar}{e^{2}/4\pi\epsilon_{0}}=\frac{\hbar_{s}}{m}
\]
the Bohr radius is given by: $a_{s}=(\hbar_{s}/m)^{2}/GM_{s}$, and
the energy of the ground state: $E_{s}=GM_{s}/2a_{s}$. We can express
these quantities in terms of the solar mass: $\hbar_{s}=kh_{\odot}$,
$a_{s}=ka_{\odot}$, and $E_{s}=kE_{\odot}$, where $k=M_{s}/M_{\odot}$
called the scale factor. The theoretical parameters are given by:
$h_{\odot}/m=7.62311\times10^{14}\mbox{ Js/kg}$, $a_{\odot}=2.92705\times10^{-2}\mbox{ AU}$
and $E_{\odot}/m=-15.1538\mbox{ GJ/kg}$.

\section{ORBITAL DYNAMICS MODEL}

In classical mechanics, the trajectory followed by a planetesimal,
orbiting a star, is an ellipse with semi-major axis $a$ and eccentricity
$\epsilon$. The average value of the distance planetesimal-star during
a complete orbit, turns out to be $\left\langle {r}\right\rangle =a(1+\epsilon^{2}/2)$.
The total mechanical energy is $E/m=-GM_{s}/2a$ and the magnitude
of the orbital angular momentum is $L/m=\sqrt{GM_{s}a(1-\epsilon^{2})}$.
By conservation of angular momentum, in the accretion process of planetesimals,
planets must be in the same plane $\theta=0$. The equation of motion
for the $i$th planet orbiting a star is: $\ddot{r_{i}}=-G(M_{s}+m_{i})/r_{i}^{2}+F_{i}$,
where $F_{i}$ are small magnitudes, which contain the perturbative
effects of all other objects on the $i$th planet, which is not a
point particle; this causes the parameters which determine the orbit,
vary periodically. The variation of the orbital kinetic energy, due
to the variation of the eccentricity is: $\triangle E_{\epsilon(i,f)}/m=GM_{s}/2\left[(1-\epsilon_{f}^{2})/r_{f}-(1-\epsilon_{i}^{2})/r_{i}\right]$
and, due to the variation of the orbital inclination: $\triangle E_{\theta(i,f)}/m=\left(2GM_{s}/r_{i}\right)sin^{2}\left[(\theta_{f}-\theta_{i})/2\right]$.

\begin{figure}
\includegraphics[scale=0.5]{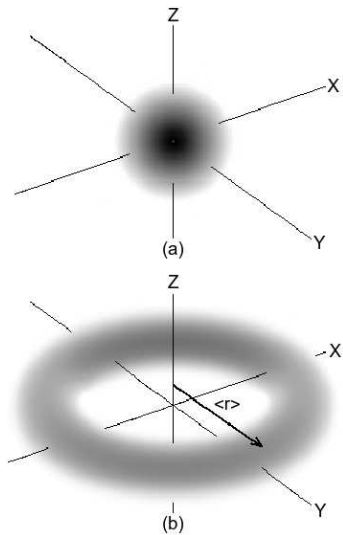} \caption{Probability of finding a planetesimal at a point in space, $\left|\rho\psi_{n}\left(\rho,\theta,\phi\right)\right|^{2}$,
for (a) $n=1$ and (b) $n>1$.}

\label{fig:Toroide} 
\end{figure}

In the quantum-like model, by substituting the Hamiltonian, $\widehat{H}=-\hbar_{s}^{2}\nabla^{2}/2m-GM_{s}m/r$
($\nabla^{2}$ is the Laplacian in spherical coordinates) in the Schrödinger-like
equation, we obtain: $E_{n}/m=-E_{s}/n^{2}$, where $E_{s}=GM_{s}/2a_{s}$
and $a_{s}=(\hbar_{s}/m)^{2}/GM_{s}$. Also, the average distance,
$\left\langle {r}_{n,\ell}\right\rangle =a_{s}/2\left[3n^{2}-\ell\left(\ell+1\right)\right]$;
the magnitude of the angular momentum, $L_{\ell}/m=\sqrt{\ell\left(\ell+1\right)}\hbar_{s}$,
and its orientation, $\theta_{\ell,m_{\ell}}=\arccos\left(m_{\ell}/\sqrt{\ell\left(\ell+1\right)}\right)$.
The quantum numbers leading to orbit and where the probability is
maximum, are $n=1,2,\ldots$, $\ell=n-1$ and $m_{\ell}=\pm\ell$.
The sign of $m_{\ell}$ corresponds to normal and retrograde orbits,
respectively. With the wave function: 
\[
\psi_{n}=\frac{2^{n}}{n^{2}\sqrt{(2n-1)!}}\left(\frac{\rho}{n}\right)^{n-1}e^{-\rho/n}Y_{n-1}^{\pm\left(n-1\right)}\left(\theta,\phi\right)
\]
(where $\rho=r/a_{s}$), the probability of finding a planetesimal
in space $\left|\rho\psi_{n}\left(\rho,\theta,\phi\right)\right|^{2}$
is obtained (see Figure \ref{fig:Toroide}). 

\noindent If the planetesimal obeys the quantum-like model, depending
on the distribution of matter and the process of accretion of planetesimals,
a large object (planet, dwarf planet or asteroid) is formed on the
mean value of the distance $\left\langle r_{t}\right\rangle \equiv a_{s}n(n+1/2)$,
ie, the object must orbit the star formed, following an elliptical
trayectory with semi-major axis $a_{t}\equiv a_{s}n^{2}$, eccentricity
$\epsilon_{t}\equiv1/\sqrt{n}$ and inclination $\theta_{t}\equiv\arccos\left|\sqrt{1-1/n}\right|$
(because this is the most probable, and stable closed orbit); with
a total energy, $E_{t}/m\equiv-E_{s}/n^{2}$ and magnitude of the
angular momentum, $L_{t}/m\equiv\sqrt{n(n-1)}\hbar_{s}$. The orbital
inclination and eccentricity of the formed objects changes cyclically
over time \citep{Lisiecki}, so the observed total energy $E_{o}/m=-GM_{s}/2a_{o}$
should be corrected: 
\begin{equation}
E_{c}=E_{o}+\triangle E_{\epsilon(t,o)}-\triangle E_{\theta(t,o)}+\triangle\varepsilon.\label{eq:Ec}
\end{equation}
in this equation $\triangle\varepsilon$ corresponds to other effects,
which generally lead to a loss of energy, such as planet-asteroids
collisions and others.

\section{THE SOLAR SYSTEM}

\begin{figure}
\includegraphics[scale=0.8]{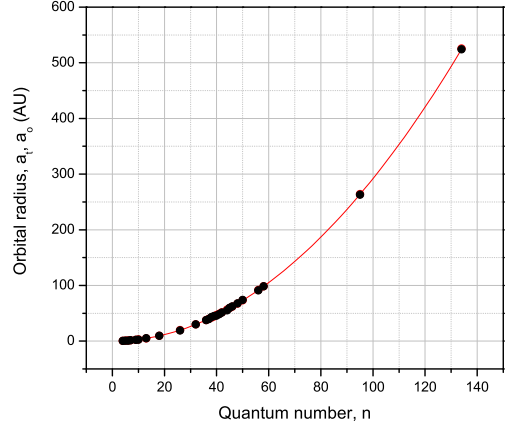} \caption{Orbital radius (planets): theoretical, $a_{t}$ (solid line) and observed,
$a_{o}$ (circles).}
\label{fig:SSRadius} 
\end{figure}

\noindent On the origins of the solar system, the formation of objects
by the process of accretion of planetesimals can occur anywhere in
the protoplanetary disk, but the stable closed orbits give rise to
larger objects, according to the quantum-like model, this occurs in
$a_{t}=ka_{\odot}n^{2}$. Based on the uncertainty principle, $\triangle p\triangle a_{t}\simeq\hbar_{s}$,
we have that $\triangle a_{t}=ka_{\odot}$. For an average radius
of $\sim300\mbox{ km}$, the icy moons and rocky asteroids in our
solar system passing from a rounded potato to a sphere \citep{Lineweaver},
for this reason, we selected objects with a diameter greater than
Vesta $\geq525.4\mbox{ km}$ (NASA JPL Small-Body Database), smaller
objects are discarded because they can easily change orbit, due to
collisions or internal dynamic processes (eg, Yarkovsky effect). Assuming
that the scaling factor is $k=1$, assigning a quantum number to each
object and selecting the of greater diameter: (4) Mercury, (5) Venus,
(6) Earth, (7) Mars, (9) Vesta, (10) Ceres, (13) Jupiter, (18) Saturn,
(26) Uranus, (32) Neptune, (36) 78799 (2002 XW93), (37) Pluto, (38)
Haumea, (39) (2010 KZ39), (40) Makemake, (41) (2013 FZ27), (42) 42301
(2001 UR163), (44) 84522 (2002 TC302), (45) (2013 FY27), (46) (2010
RE64), (48) Eris, (50) 229762 (2007 UK126), (56) 145451 (2005 RM43),
(58) (2008 ST291), (95) (2012 VP113), (134) Sedna, (Figure \ref{fig:SSRadius}).
The filling of the orbits depends mainly on the mass of the star and
the amount of material available in the protoplanetary disk. The Bohr
radius value observed is obtained: $a_{\odot}^{o}=\left(2.92007\pm0.00256\right)\times10^{-2}\mbox{ AU}$,
$\chi_{DoF}^{2}=0.30574$, $R^{2}=0.99997$, and percentage error
$e\%=0.24$ with respect to the theoretical value.

\noindent 

\begin{figure}
\includegraphics[scale=0.8]{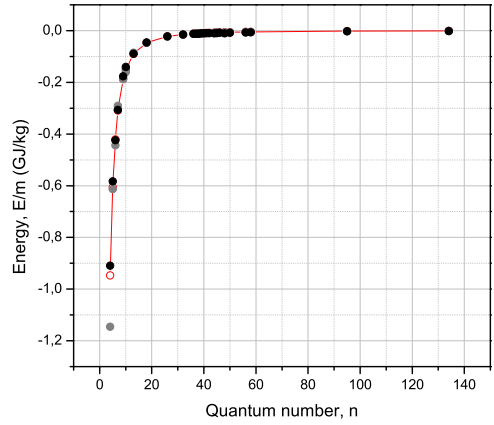} \caption{Energy (planets): theoretical $E_{t}$ (solid line), observed $E_{o}$
(gray circles), and corrected $E_{c}$ (black circles).}
\label{fig:SSEnergy} 
\end{figure}

Figure \ref{fig:SSEnergy} shows the relationship between the corrected
energy $E_{c}$, equation (\ref{eq:Ec}), and the quantum numbers.
Once have formed interior objects orbital radius is maintained, while
objects in the energy continuum ($\geq30\mbox{ AU}$) can be easily
moved (Kuiper belt). Jupiter is at the inflection point. Is obtained
$E_{\odot}^{o}/m=(14.65188\pm0.00160)\mbox{ GJ/kg}$, $\chi_{DoF}^{2}=0.00002$,
$R^{2}=0.99956$, and $e\%=3.31$. It is assumed that the percentage
error is due mainly to the effect they have had the impact of asteroids
on the inner solar system objects. In turn, the degree of fit, is
evidence of the quantization of angular momentum, ie, the quantization
of the orbital inclination and magnitude of the angular momentum in
the period of formation of the solar system.

\begin{figure}
\includegraphics[scale=0.8]{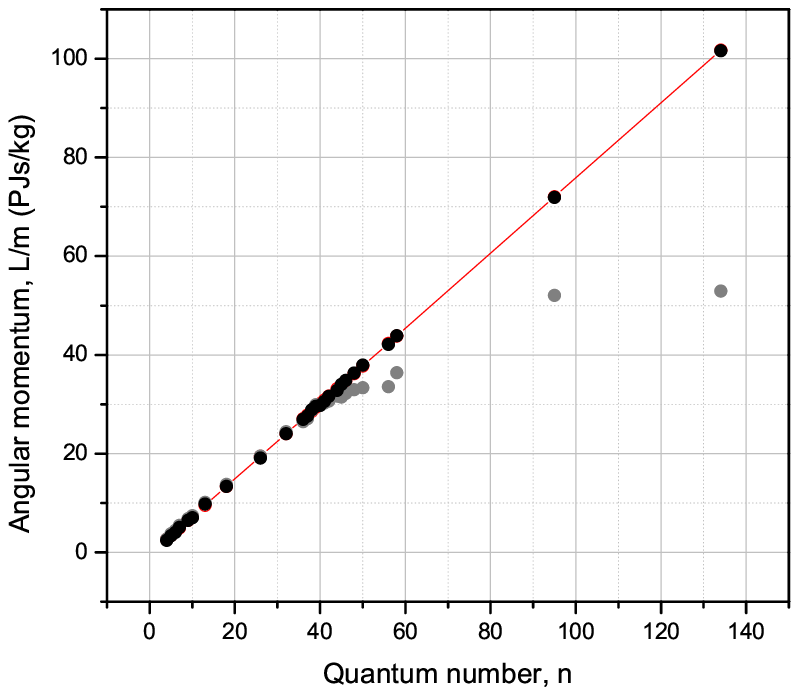} \caption{Orbital angular momentum (planets): theoretical $L_{t}$ (solid line),
observed $L_{o}$ (gray circles), and corrected $L_{c}$ (black circles).}
\label{fig:SSMAngular} 
\end{figure}

In Figure \ref{fig:SSMAngular} shows the observed angular momentum,
which depends on the eccentricity, $L_{o}/m=\sqrt{GM_{s}a_{o}\left(1-\epsilon_{o}^{2}\right)}$;
the objects far away from the Sun have very high orbital inclinations,
this generates a variation in the orbital eccentricity (\ref{eq:Ec}),
which in turn produces an anomalous behavior in the angular momentum.
Suppressing effect of orbital inclination on the eccentricity, we
can obtain a corrected angular momentum: $L_{c}/m=\sqrt{GM_{s}r_{o}\left(1-\epsilon_{c}^{2}\right)}$,
is obtained: $\hbar_{\odot}^{o}/m=(0.76144\pm0.00081)\mbox{ PJs/kg}$,
$\chi_{DoF}^{2}=0.03682$, $R^{2}=0.99992$, and $e\mbox{\%}=0.11$.

The Sun corresponds the quantum number $n=1$, consequently, its angular
momentum is zero and the probability distribution of matter is a sphere,
see Figure \ref{fig:Toroide}(a), this could explain why the angular
momentum of the Sun corresponds only $\sim2\mbox{\%}$ of the entire
solar system. The quantum numbers $n=2,3,\ldots$, have a probability
distribution of matter in the form of concentric toroids, see Figure
\ref{fig:Toroide}(b), which have different degrees of inclination
and orbital velocity, the absence of planets in numbers quantum $n=2,3$,
indicates that objects migrated outwards or planetesimals were captured
by the Sun.

\subsection{Orbital migration}

The parameters obtained correspond to the current solar system, but
the planets had a different distribution when they formed: it is assumed
that the sequence of the inner planets was, (4) Mercury, (5) Venus,
(6) Earth, (7) Mars, (8) Vesta and (9) Ceres. The Jovian planets has
the sequence given by \citep{Gomes}. In this solar system objects
representing at least 26 is considered that at the time of its formation,
the same objects were present. Taking the maximum goodness of fit
(maximum likelihood) for these conditions we obtain the sequence:
(4) Mercury, (5) Venus, (6) Earth, (7) Mars, (8) Vesta, (9) Ceres,
(13) Jupiter, (16) Saturn, (19) Neptune (21) Uranus, (34) 78799 (2002
XW93), (35) Pluto, (36) Haumea, (37) Makemake, (38) 55565 (2002 AW197),
(39) (2010 RF43), (40) 42301 (2001 UR163), (41) 84522 (2002 TC302),
(42) (2004 XR190), (43) (2013 FY27), (44) (2010 RE64), (45) 225088
(2007 OR10), (46) Eris, (48) 229762 (2007 UK126), (53) 145451 (2005
RM43), (55) (2008 ST291), (90) (2012 VP113), (127) Sedna and $a_{\odot}^{o}=(3.25085\pm0.00279)\times10^{-2}\mbox{ AU}$,
$\chi_{DoF}^{2}=0.29805$, $R^{2}=0.99997$, and a scale factor $k=1.11062$,
therefore the mass of the sun was $M_{s}\simeq1.1M_{\odot}$, which
agrees with \citep{Boothroyd}, and \citep{Guzik}.

\subsection{Satellites}

\begin{figure}
\includegraphics[scale=0.8]{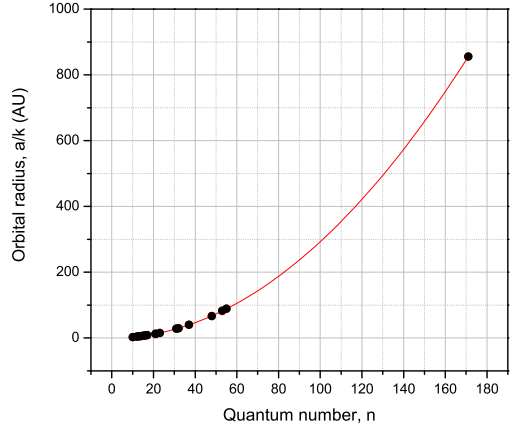} \caption{Orbital radius (satellites): theoretical $a_{t}$ (solid line) and
observed $a_{o}$ (black circles).}
\label{fig:SARadius} 
\end{figure}

Is have selected the satellites of the planets of the solar system
(Planetary Satellite Physical Parameters, JPL, NASA) with diameters
$\geq350\mbox{ km}$. In order to compare with the solar system, has
taken the ratio of the orbital radius and the scaling factor ($a_{o}/k$).
By the same procedure, used for the solar system, is obtained: (10:J)
Io, (12:S) Mimas, (13:J) Europe, (14:S) Enceladus, (15:S) Tethys,
(16:J) Ganymede, (17:S) Dione, (21:S) Rhea, (21:J) Callisto, (23:N)
Proteus, (31:S) Titan, (32U) Ariel (37:U) Umbriel, (48:U) Titania,
(53:S) Iapetus, (55:U) Oberon, (171:E) Moon. Was excluded (40:N) Triton
because anomalous behavior with respect to the model, shows that is
a captured by Neptune object. The values ??obtained for the parameters
are observed: $a_{\odot}^{o}=(2.92645\pm0.00173)\times10^{-2}\mbox{ AU}$
($\chi_{DoF}^{2}=0.26477$, $R^{2}=0.99999$, and $e\%=0.02$ (see
Figure \ref{fig:SARadius}). Is obtained: $\hbar_{\odot}^{o}/m=(0.76581\pm0.00185)\mbox{ PJs/kg}$,
$\chi_{DoF}^{2}=0.14702$, $R^{2}=0.99982$, and $e\%=0.46$; $E_{\odot}^{o}/m=(15.11796\pm0.02155)\mbox{ GJ/kg}$,
$\chi_{DoF}^{2}=1.268\times10^{-7}$, $R^{2}=0.99993$, and $e\%=0.24$.

\section{EXOPLANETS}

\begin{figure}
\includegraphics[scale=0.8]{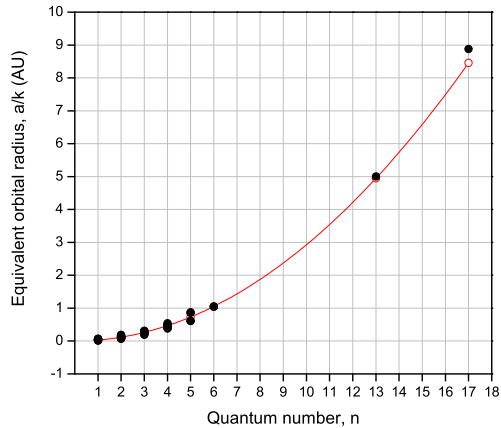} \caption{Orbital radius (exoplanets): theoretical $a_{t}$ (solid line) and
observed $a_{o}$ (black circles).}
\label{fig:EXRadius} 
\end{figure}

This research has made using the Exoplanet Orbit Database and the
Exoplanet Data Explorer at exoplanets.org \citep{Wright}. Have been
selected systems with a single star with known mass and exoplanets,
radio and orbital eccentricity known; assigning a quantum number and
selecting the most massive, there are 219 exoplanets. In order to
compare with the solar system, has taken the ratio of the orbital
radius and the scaling factor ($a_{o}/k$). The value obtained for
the Bohr radius is: $a_{\odot}^{o}=(3.03542\pm0.00944)\times10^{-2}\mbox{ AU}$,
$\chi_{DoF}^{2}=0.00108$, $R^{2}=0.99777$, and $e\mbox{\%}=3.70$
(see Figure \ref{fig:EXRadius}).

\begin{figure}
\includegraphics[scale=0.8]{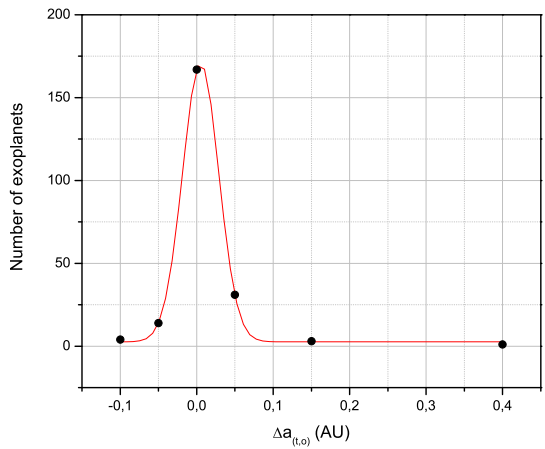} \caption{Number of exoplanets vs. difference between the orbital radius theoretical
$a_{t}$, and observed $a_{o}$.}
\label{fig:EXFrequency} 
\end{figure}

In the Figure \ref{fig:EXFrequency} is shown the number of exoplanets
vs. difference between the observed and theoretical orbital radius
($\triangle a/k$), has made an adjustment to a Gaussian distribution
$y=y_{0}+(A/w\sqrt{\pi/2})\exp\left[-2(\triangle a/k\, w)^{2}\right]$,
here $y_{0}=2.66240\pm0.88034$, $A=10.01864\pm0.20650$, $w=0.04750\pm0.00094$,
$\overline{a/k}=0.00517\pm0.00072$, $\chi_{DoF}^{2}=2.32095$, and
$R^{2}=0.99978$. This means that exoplanets are not formed anywhere,
only preferentially in the orbits given by $a_{t}/k=a_{\odot}n^{2}$.

In classical mechanics, by conservation of angular momentum, the objects
must be in the same plane, $\theta\rightarrow0^{\circ}$ (in the accretion
disk), in the quantum-like model there is a quantization of angular
momentum and its orientation, objects have higher orbital inclination
between the closer you are to the star, this explains the elevated
orbital inclinations found for exoplanets ($\theta\rightarrow90^{\circ}$).
It have $h_{\odot}^{o}/m=(0.76619\pm0.002910\mbox{ PJs/kg}$, $\chi_{DoF}^{2}=0.00681$,
$R^{2}=0.99638$, and $e\mbox{\%}=0.51$.

\begin{figure}
\includegraphics[scale=0.8]{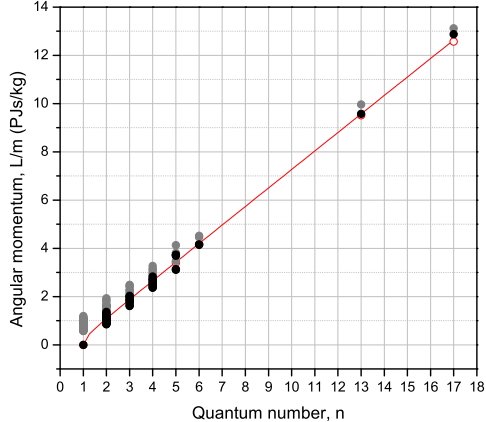} \caption{Orbital angular momentum (exoplanets): theoretical $L_{t}$ (solid
line), observed $L_{o}$ (gray circles), and corrected $L_{c}$ (black
circles).}
\label{fig:EXMAngular} 
\end{figure}

The total energy shown large variations when the exoplanet is very
close to the star (before continuous), this is due to variations in
the eccentricity ($\triangle E_{\epsilon(t,o)}$) and the inclination
($\triangle E_{\theta(t,o)}$) of the orbit, is obtained $E_{\odot}^{o}/m=(17,57860\pm0.36720)\mbox{ GJ/kg}$,
$\chi_{DoF}^{2}=0.24013$, $R^{2}=0.88301$, and $e\mbox{\%}=16.00$.

\subsection{Protoplanetary disk around HL Tauri}

Considering a central protostellar mass for HL Tau given by $k=0.55$
\citep{Beckwith,Sargent}, has to be formed planets in $a_{t}=ka_{\odot}n^{2}$.
Based on uncalibrated image of the protoplanetary disk HL Tauri taken
by ALMA (ESO / NAOJ / NRAO) where planets may be forming, several
gaps are distinguished $a_{t}=(10.06,\,24.49,\,46.94,\,68.02)\mbox{ AU}$
which correspond to $n=(25,\,39,\,54,\,65)$, respectively.

\section{CONCLUSIONS}

The quantum-like model is applicable to any system that quantize the
action, therefore, the formalism of quantum mechanics can be used
in macroscopic systems where, the quantum of action, $\hbar_{s}\gg\hbar$.
In this particular case, this occurs because the flow of gravitational
field is quantized. The dynamics is determined only by the mass of
the central object (as the case, star or planet) independently of
the mass of the orbiting object. The probability distribution of planetesimals
in space and the physical quantities (radius, eccentricity, energy,
orbital angular momentum and its inclination) of the objects in the
planetary system, take a certain value (quantized). Therefore, planets
and exoplanets tend to be formed at a predetermined distance which
depends on the mass of the star and the quantity of matter on the
disc, obeying $a_{t}=ka_{\odot}n^{2}$; the orbit may be normal or
retrograde. Similarly, the angular momentum and its orientation takes
certain value, however, because the planets and exoplanets are not
punctual objects, occur periodic variations, in the eccentricity and
inclination orbital, causing a change in the value of the observed
energy, equation (\ref{eq:Ec}). The obtained theoretical model can
be used as a tool to parameterize and study the formation of planetary
systems.

\section*{ACKNOWLEDGMENTS}

We thank the Dirección de Investigaciones (DIN) of the Universidad
Pedagógica y Tecnológica de Colombia (UPTC).

\appendix

\section{QUANTUM-LIKE MODEL}

In the protoplanetary disk particles condense into small planetesimals
of mass $m$, which orbit the star. Since the mass of the star, $M_{s}$,
is extremely large compared to the mass of planetesimals, $M_{s}\gg m$,
action star prevails and the interaction between them, only results
in weak perturbations to their respective orbits. The Hamiltonian
(in polar coordinates) for a planetesimal orbiting a star is: $H=p_{r}^{2}/2m+U(r)$,
where $p_{r}=m\dot{r}$ and $U(r)$ is an effective potential energy
$U(r)=L_{o}^{2}/2mr^{2}-GM_{s}m/r$, in turn $L_{o}\equiv p_{\theta}=mr^{2}\dot{\theta}$
is the angular momentum. As the force is conservative, $F=-\nabla U$;
planetesimal experience a attractive force, for $r>r_{o}$, and repulsive
force, for $r<r_{o}$, while $r=r_{0}$ forces cancel each other out.
It can easily be shown that $U(r)$ has a minimum at $r=r_{o}$: $U_{o}=U(r)_{r=r_{o}}$
necessarily $\left(\nabla U\right)_{r=r_{o}}=0$ and $k_{o}=\left(\nabla^{2}U\right)_{r=r_{o}}>0$.

The lower energy state corresponds to $E=U_{o}$, planetesimal path
is a circumference of radius $r_{o}$ centered on the star. The planetesimal
on an orbital cycle runs its perimeter $(x_{o}=2\pi r_{o})$ at a
time ($t_{o}$). As, $\nabla U_{o}=0$ the planetesimal in this kind
of orbit equivalently behaves a free particle, in this case the equation
of Hamilton-Jacobi is given by: $\left(\nabla S\right)^{2}/2m+\partial S/\partial t=0$.
Is denoted by $x$ the perimeter distance traveled by the planetesimal.
As in the Hamiltonian coordinate $x$ (cyclic), or the time $t$ does
not appear explicitly, there are two conserved quantities: the linear
momentum ($\nabla S\equiv p_{o}$), and the total energy of the system
($-\partial S/\partial t\equiv E_{o}$). Consequently the action is
separable: $S(x,t)=S_{x}(x)+S_{t}(t)=p_{o}x-E_{o}t$, and is periodic:
$S(x,t)=S(x+x_{o},t+t_{o})$.

Bertrand's theorem shows that under any initial condition, the only
potential producing stable orbits have the functional form $\sim1/r$
(Kepler) or $\sim r^{2}$ (harmonic oscillator); additionally closes
the orbit if the relationship between the radial and orbital frequency
is $\omega/\omega_{o}=n/m$, where $n$ and $m$ relatively prime.
By disturbing the orbit of a planetesimal it can be shown that the
planetesimal oscillates around its original path. A perturbed orbit
if after $n$ periods of variation of $r$ ($t_{o}=n\tau$), the complete
revolutions planetesimal $m$, $n\triangle\theta=m2\pi$ is repeated
and a revolution in the planetesimal travels a distance $\lambda=\triangle\theta r_{o}$,
where Bohr's rule is obtained for the orbits, $n\lambda=2\pi r_{o}$.

Hamilton's principal function vanishes for: $S(0,0)=S(x_{o},t_{o})=n(p_{o}\lambda)-n(E_{o}\tau)=0$.
The terms inside the parentheses correspond to a constant which is
denoted by $h_{s}$. Note that the action is quantized, that is, $S_{x}(x_{o})=S_{t}(t_{o})$
is given by $n$ times a quantum of action $h_{s}$, we will call
macroscopic parameter, which must be determined for this physical
system. The quantum of action $h_{s}$ when the planetesimal travels
a $x_{o}=\lambda$ is $h_{s}=S_{x}(\lambda)$, or when time elapses
$t_{o}=\tau$ is, $h_{s}=S_{t}(\tau)$; these expressions are analogous
to the relations of de Broglie and Planck, respectively:

\begin{equation}
p_{o}=\hbar_{s}k,\qquad E_{o}=\hbar_{s}\omega,\label{eq:Broglie-Planck}
\end{equation}

\noindent where, $\hbar_{s}=h_{s}/2\pi$ called, reduced macroscopic
parameter, $k=2\pi/\lambda$ and $\omega=2\pi/\tau$.

The perturbation of the trajectory of a planetesimal is caused by
the presence of other planetesimals, causing it to move from its original
path $r_{o}$, a distance $\xi=r-r_{o}$ (which is very small, $r\simeq r_{o}$).
Making a Taylor series expansion of the effective potential $U(r)$
around of $r_{o}$, its radial motion is given by: $p_{\xi}^{2}/2m+k_{o}\xi^{2}/2=E-U_{o}$,
where $p_{\xi}=m\dot{\xi}$ is a transverse momentum with respect
to the unperturbed motion. Consequently, the disturbance makes the
planetesimal harmonically oscillate around the non-perturbed trajectory.

Based in Hamilton canonical equation, $\dot{p_{\xi}}=-\partial H/\partial\xi$,
we may obtain the wave equation, $\partial^{2}\psi(x,t)/\partial t^{2}-v_{o}^{2}\nabla^{2}\psi(x,t)=0$;
where $\psi(x,t)=\xi(x)\xi(t)$ is defined and $v_{o}=\omega/k$.
The corresponding solution has the form: $\psi=C\,\exp\left[i(k\, x-\omega\, t)\right]$.
As the path is periodic, $\psi(0,0)=\psi(x_{o},t_{o})=0$ then: $k\rightarrow k_{n}=n\pi/\lambda$,
$\omega\rightarrow\omega_{n}=n\pi/\tau$ being ($n=1,2,\ldots$).
As we can see, the boundary conditions have the effect of discretizing
the wave number and frequency. The solution of the wave equation takes
the form: $\psi_{n}(x,t)=C_{n}\,\exp\left[i(k_{n}x-\omega_{n}t)\right]$.

However, substituting relations (\ref{eq:Broglie-Planck}) in $\psi_{n}(x,t)$
there is no correspondence between the wave speed and the speed of
planetesimals. Planetesimals to represent using the wave equation,
it is necessary to modulate the amplitude of the wave with a positive
real function, $A(x,t)=A(x-v_{o}t)$, to limit the extent of the wave
and cause the envelope to move with a group velocity, $v_{o}$. Obtaining,
in general:

\begin{equation}
\Psi\left(r,t\right)=C\, A(r,t)\,\exp\left[\frac{i}{\hbar_{s}}S(r,t)\right],\label{eq:wave-function}
\end{equation}

\noindent the exponential allows taking into account the phenomena
related to the superposition of waves, while the coefficient corresponds
to a distribution function (or probability density) to find the planetesimal
in space: $\varrho=\left|\Psi(r,t)\right|^{2}$. By integrating over
all space the complex constant $C$ can normalize to unity. The probability
current is defined as $J=\varrho v_{o}$. Conservation of probability
is given by the continuity equation:

\begin{equation}
\frac{\partial\varrho}{\partial t}+\nabla J=0,\label{eq:continuity-equation}
\end{equation}

\noindent the probability density moves in space with the same velocity
$\nabla S/m=v_{o}$ and follow the path set of planetesimals. The
Hamilton-Jacobi equation we add the term $U_{Q}=\frac{1}{2}k_{o}\xi^{2}$
which corresponds to the effect of the disturbance,

\begin{equation}
H+U_{Q}+\frac{\partial S}{\partial t}=0,\label{eq:Hamilton-Jacobi-Q}
\end{equation}

\noindent this implies that the classical force is being affected
by another force, which generates the quantization of the action of
the physical system: $F=-\nabla U-\nabla U_{Q}$. The eikonal equation:
$\nabla^{2}A+(2mU_{Q}/\hbar_{s}^{2})A=0$ allows us to define a relationship
between $U_{Q}$ and $A$. Substituting this expression in equation
(\ref{eq:Hamilton-Jacobi-Q}) and with (\ref{eq:continuity-equation})
we can construct an analogous equation to Schrödinger, which call
Schrödinger-like:

\begin{equation}
H\Psi\left(r,t\right)=i\hbar_{s}\frac{\partial}{\partial t}\Psi\left(r,t\right).\label{eq:Schrodinger}
\end{equation}

From the Hamilton-Jacobi equation (\ref{eq:Hamilton-Jacobi-Q}) we
obtain the trajectory followed by the particles and from Schrödinger
equation (\ref{eq:Schrodinger}) we obtain the distribution of particles
in the space or the probability density, $\rho(r,t)$.

In the star-planetesimal system planetesimals that quantized action
giving rise to stationary waves (or stable trajectories) which store
the energy of the disturbance. These waves interfere constructively
giving rise to phenomena of resonance or stationary states (when the
system with the resonance frequency is disturbed), are described by
the eigenvalue equation $H\varphi\left(r\right)=E_{n}\varphi\left(r\right)$.
Planetesimals not quantized action are represented by traveling waves
transporting the disturbance energy, these waves end transferring
its energy, so that the planetisimals assume the classical trajectory,
which is described by (\ref{eq:Hamilton-Jacobi-Q}) where $U_{Q}=0$,
ie, the formalism of classical mechanics. Therefore, we have a privileged
system where states appear when $U_{Q}\neq0$, which can be described
with the formalism of quantum mechanics. 
\end{document}